\documentclass{aa}
\usepackage{graphics}

\newcommand{\Mbol}{$M_{\mathrm{bol}}$}
\begin{document}

   \title{AGB stars in the Magellanic Clouds. I. The C/M ratio.}
   
   \author{M.-R.L. Cioni\inst{1}
          \and
           H.J. Habing\inst{2}
          }
 
   \offprints{mcioni@eso.org}

   \institute{European Southern Observatory,    
              Karl--Schwarzschild--Stra\ss e 2, 
              D--85748 Garching bei M\"{u}nchen, Germany
              \and
              Sterrewacht Leiden,
              Niels Bohrweg 2,
              2333 RA Leiden, The Netherlands
             }

   \date{Received .../ Accepted ...}

   \titlerunning{The C/M ratio in the Magellanic Clouds}
  
   \authorrunning{Cioni \& Habing}
   
   \abstract{Regions of different  metallicity have been identified in
     the Magellanic Clouds by using the ratio between Asymptotic Giant
     Branch stars of  spectral type C and M.   In the Large Magellanic
     Cloud the ratio  appears to decrease radially while  in the Small
     Magellanic Cloud (SMC) there is no clear trend,  reflecting either the
     large extension  of the  SMC along  the line of  sight or  a more
     complex star formation history. The distribution of the C/M ratio
     is clumpy and corresponds to a  spread in [Fe/H] of $0.75$ dex in
     both Clouds. There is an indication of increasing C/M ratio, thus
     decreasing metallicity, towards  the Bridge region connecting the
     two Clouds.}

\maketitle

\section{Introduction}

Asymptotic Giant Branch (AGB) stars trace the intermediate age
population of galaxies (ages between 1 and several Gyr).  In
near--infrared (near--IR) light they are often the brightest isolated
objects in a galaxy and can be stu\-di\-ed at distances beyond 1 Mpc.
There are two main kinds of AGB stars: oxygen--rich stars (more
O--atoms than C-atoms, spectral type M or K) and carbon--rich stars
(more C--atoms than O---atoms, spectral type N or C). When the number of
O--atoms equals that of C--atoms the AGB star is of type S. It is not
straight forward to identify S stars either photometrically or
spectroscopically, and this introduces as well intermediate classes of
AGB stars (e.g. MS and SC). Single stars enter the AGB phase as
oxygen--rich but may be converted into carbon--rich stars after
several short episodes in which matter that has been enriched in
carbon nuclei by nuclear fusion is dredged--up to the stellar surface;
these episodes are called thermal pulses; stars that have become
C--stars in this way are called ``intrinsic'' C--stars; they may all
have N--type spectra.

Searches for carbon  and M--type stars in Baade's  Window (Glass et al.
\cite{glass99}), in the solar neighbourhood (Mikami \cite{mikami}) and
in  the Magellanic  Clouds (i.e.  Blanco et  al.~\cite{bmb},  Feast \&
Whitelock \cite{feast}) have shown that  the ratio of number of carbon
stars  to that  of  oxygen--rich  AGB stars  varies  strongly from  one
environment to another and that  it correlates with metallicity in the
sense:  lower metallicity, more  carbon stars.   For example,  Cook et
al. \cite{cook} (and references therein)  derived a C/M ratio in seven
gas--rich galaxies  and found that this correlates  very well inversely
with the  abundance ratio log[O/H]: higher abundance  lower C/M ratio --
see their figure 6.

Iben  \&   Renzini  (\cite{iben})  explain   the  correlation  between
metallicity  and  the  C/M  ratio  by three  factors:  (1)  for  lower
metallicity the  O--rich stars  will turn into  C--type AGB  stars more
easily and  stay so for a longer  time on the AGB  because less carbon
atoms  need to  be dredged--up  for the  transformation; (2)  at lower
metallicity  the evolutionary  tracks  for AGB  stars  move to  higher
temperatures  and  the  number  of  M5+\footnote{When  the  sample  of
M-giants consists  of stars  with spectral type  M0 and later  we will
designate this as ``M0+''; similarly  we will use M3+, M5+ etc.} stars
drops considerably;  the abundance of  TiO molecules is lower  and the
atmosphere  is  more   transparent;   (3)  in   very--low metallicity
environments a  post--horizontal--branch star  may become a  white dwarf
without becoming first an AGB star (``AGB--manqu{\'e} stars'').

In this article we use the DCMC, the DENIS Catalogue of point sources
detected in the $IJK_s$--photometric bands in the direction of the
Magellanic Clouds (Cioni et al. \cite{dcmc}). We eliminate foreground
stars, we select red stars with a luminosity above the tip of the RGB
(=TRGB) and we use photometric colours to distinguish between C--type
AGB stars and AGB stars of class M0+.  We then study the variation of
the C/M0+ ratio over the face of each Magellanic Cloud.  We exploit
two advantages of the DCMC catalogue over earlier studies: full
coverage of both Clouds ($320$ square degrees in the LMC and $150$ in
the SMC) and the inclusion of M giants of all M-type spectra (M0+).

By considering only AGB stars above the TRGB we eliminate stars with a
carbon--type spectrum, but NOT on the AGB: in a binary in which the
secondary is a white dwarf, the primary may have a C--rich spectrum
because it received enough C--enriched material from the secondary
when, in the past, that was an AGB star.  Such stars are called
``extrinsic C--stars''; a thorough discussion of their properties is
given by Jorissen (\cite{jorissen}). All R-type carbon stars,
CH--stars and Ba--stars seem to be such extrinsic stars.  Carbon stars
with a luminosity above that of the TRGB must be AGB stars and thus
very probably intrinsic carbon stars.

In Sect. $2$ we discuss the sample and our selection criteria and we
present the C/M ratio and compare it with the values obtained during
previous researches.  Sect. $3$ addresses the metallicity distribution
derived in other papers and we compare the Magellanic Clouds
(MC) with other galaxies.  Conclusions are given in Sect. $4$.

\section{The distribution of M-type and C-type AGB stars in LMC and SMC}

\subsection{Selection of AGB stars from the DENIS catalogue}

We selected AGB stars from the DCMC survey following the same criteria
as used in an earlier paper (Cioni et al., \cite{cionimorf}).
Briefly: in each Cloud we select all stars that (1) have been detected
in all three DENIS wavelength bands ($IJK_{\mathrm s}$); that (2) have
an $I$--band magnitude above that of the TRGB ($I=14.54$ in the LMC and
$I=14.95$ in the SMC) and (3) that have a red colour $(I-J)>
(19.78-I)/4.64$ separating MC stars from foreground objects. The same
sample has been used in an earlier paper to discuss the morphology of
each Cloud (Cioni et al., \cite{cionimorf}; see region B in their
figure 1). The two samples contain almost all AGB stars above the
TRGB; we only miss stars with a circumstellar shell thicker than
$(J-K_{\mathrm s})> 2.4$. The evolution of stars with a thick
circumstellar envelope is so fast that there are always very few of
them.  Experience with ISO programmes confirms that the number of
stars with such thick shells is too small to affect our conclusions
(Loup \cite{cecile}).  The LMC sample contains 32801 AGB stars; the
SMC sample contains 7652 stars.

\subsection{Contamination by foreground stars}
Our selection criterium rejects most of the foreground stars (Cioni et
al. \cite{cionimorf}.) Does the remaining fraction of foreground stars
affects our conclusion?  In this paper Fig~\ref{colcol} contains a faint
extension to higher $(I-J)$ values at $(J-K_s)=0.9$; from DENIS 
colour--colour diagrams of stars outside  of the Magellanic  Clouds 
it follows that these are as well foreground objects. Nevertheless, this 
plume of objects contains a small number of genuine Magellanic Cloud 
stars: faint  super-giants of  spectral  type  K  or M  (Nikolaev  \&
Weinberg \cite{nw}). Foreground stars, if uniformly
distributed, will affect the ratio, but will not introduce variations
in the ratio.  Blanco \& McCarthy (\cite{bm}) observed several fields
covering a total of about $6$ and $7$ square degrees in the LMC and
SMC, respectively (see Sect.~3.2), and sampled foreground areas at the
same la\-ti\-tu\-de.  They found on average $10$ M2 foreground stars per
field ($0.12$ deg$^2$).  If the distribution of foreground stars is
the same as shows the gray scale in Fig.~2 in Cioni et al.
(\cite{cionimorf}), the enhancement towards the
East is also the distribution of early M--type foreground stars, we
would expect to find a smaller ratio in this direction than towards
that directly opposite.  But looking at Fig. \ref{lmcratio} we do not
find a noteworthy difference between the two regions. Thus we conclude
that the distribution of foreground stars does not bias the C/M ratio
that we derive.  The C/M ratios, on the other hand, are lower limits,
because there may be a foreground contamination by M dwarfs (Cook et
al.~\cite{cook}).

\subsection{Separating the M-type stars from the C-type stars}

\begin{figure}
\resizebox{\hsize}{!}{\includegraphics{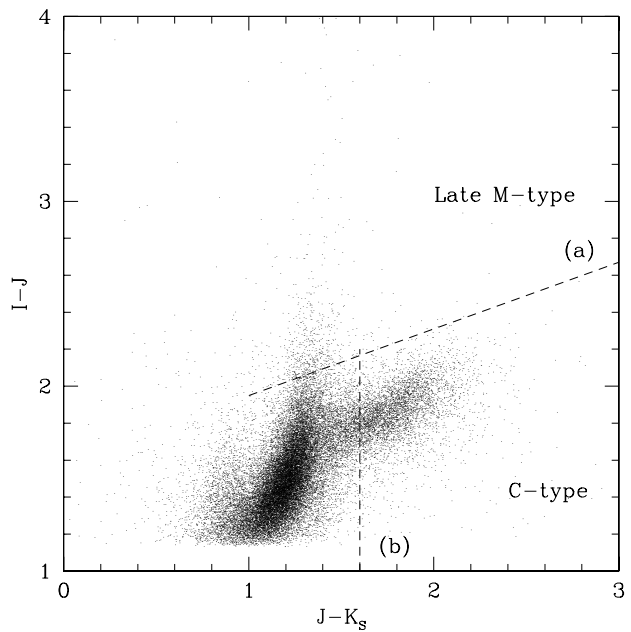}}
\caption{DENIS colour--colour diagram of AGB stars in the LMC. Dashed 
  line indicate the regions where upper M--type AGB and C--type AGB
  stars can be discriminated according to Loup et al. (\cite{loup}).}
\label{colcol}
\end{figure}

\begin{figure}
\resizebox{\hsize}{!}{\includegraphics{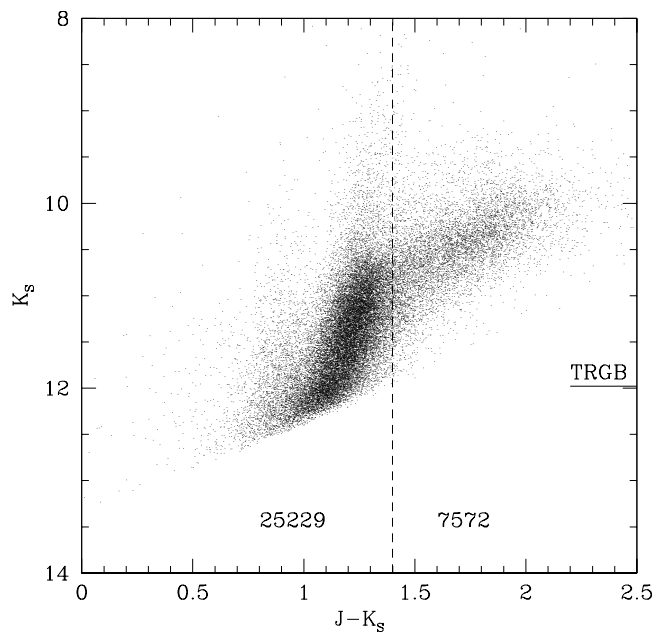}}
\caption{Colour--magnitude diagram of stars in the LMC selected to 
  be AGB stars above the TRGB ($I=14.54$, Cioni et al. \cite{cionitip}) and
  $I>-4.64\times(I-J)+19.78$ (Cioni et al. \cite{cionimorf}). The 7572
  stars to the right of the dashed vertical line are carbon stars,
  whereas the 25,229 stars to the left are M-type AGB stars.}
\label{agb}
\end{figure}

Fig.~\ref{colcol} shows all stars of the sample in a colour--colour
diagram.  A comparison one--by--one of a subsample of DCMC stars
coinciding with stars with a spectral type derived from spectroscopy
(Loup et al. in preparation) shows that the stars above the slanted
dashed line (a) are all M--type giants of spectral type M7+ and the
stars below this dashed line and to the right of the vertical line
$(J-K_{\mathrm{s}})= 1.6$ (b) are all C-stars.  This leaves the stars
in the lower left part of the diagram below line (a) and left of line
(b).  We assume that we can separate the stars in M--type stars and in
C--type stars in the LMC by the criterium ($J-K_{\mathrm{s}}$) $<$ or
$> 1.4$ (see Fig. \ref{agb}) and in the SMC $(J-K_{\mathrm{s}})=1.3$
will be the dividing line. We also assume that that the
$(J-K_{\mathrm{s}})$--colour is practically independent of spectral
type whereas the $(I-J)$--colour is correlated strongly with the
spectral type of the M--stars. This, for example has been found by
Fluks et al. (\cite{fluks}): for M0--giants $(I-J)=1.14$ and
$(J-K_{\mathrm{s}}) =1.03$ and for M9--stars these values are 4.06 and
1.25. Fluks' colours are based on stars in the solar neighbourhood.
Therefore we calibrated our relation between spectral type and the
colour $(I-J)$ on the results obtained by Blanco et al., Blanco \&
McCarthy (\cite{bmb, bm}) in the LMC and we correct the value of
$(I-J)_0$ to $1.02$ for M0--giants and to $1.99$ for M7-giants. 

The location of the C--star branch has been explained recently by
accounting properly the molecular opacity in the stellar atmosphere
when a star develops from an M--type star into a C--star (Marigo,
Girardi \& Chiosi, accepted for A\&A). We conclude that a vertical
line at $(J-K_{\mathrm{s}})=1.4$ (the dashed line in Fig. \ref{agb})
separates the M--type stars from the C-stars. The next conclusion is
that in the area of the M--giants (left of the line ($J-K_{\mathrm{s}})=
1.4$) the $(I-J)$ value determines the M--type subclass: at
$(I-J)=1.2$ the stars will be of spectral type M0; at $(I-J)=3.8$ they
will be M9+.  Close to the point $(I-J, J-K_{\mathrm{s}})=(1.4, 1.3)$
some C--stars will be mixed with the M--stars, but their fraction seems
to be quite low. S--type AGB stars have near--IR colours similar to
O--rich AGB stars and they cannot be disentangled in our diagrams; but
there are always few around.

We  apply  this  result   in  Fig.  \ref{agb},  the  ($K_S$,  $J-K_S$)
colour--magnitude diagram of our LMC  sample of AGB stars and obtain a
sample of 25229  M--type stars and of 7572  C--stars. As remarked above,
the first sample contains a small  fraction of C--stars. For the SMC we
obtain  6009  M--type  and  1643  C-type AGB  stars.   The  photometric
accuracy of DENIS at the AGB magnitudes (i.e. for \Mbol$<-4$) is below
$0.05$ mag and  the number of sources that have been  put in the wrong
section of Fig. \ref{agb} by photometric errors is negligible.

\begin{table}
\caption{C/M ratio in the Magellanic Clouds}
\label{tabij}
\[
\begin{array}{lccccc}
\hline
\noalign{\smallskip}
\mathrm{Cloud} & \mathrm{C} & \mathrm{C/M0+} & \mathrm{C/M3+} 
& \mathrm{C/M5+} & \mathrm{C/M7+}\\
\mathrm{LMC} & 7572 & 0.30 & 0.47 & 1.90 & 25.1 \\
\mathrm{SMC} & 1643 & 0.27 & 0.87 & 9.28 & 48.3 \\
\noalign{\smallskip}
\hline
\end{array}
\]
\end{table} 

Using the $(I-J)$ colours of M giants as indicated by Blanco et al.
(\cite{bmb}) and correcting for extinction ($A_{I-J}=0.16$ for
$E(B-V)=0.15$ as in Cioni et al. \cite{cionitip}) we obtain the
numbers of C-stars and the numbers of M0+, M3+, M5+ and M7+ stars.
The numbers of giants of C and M spectral subtype with the
corresponding ratio are given in, or can be easily obtained from,
table \ref{tabij}. The table shows the population of early M--type AGB
stars. The C/M0+ numbers are very similar for both LMC and SMC, which
may indicate a similar average metallicity for both Clouds.

\subsection{Total number of carbon stars}
Blanco \& McCarthy  (\cite{bm}) expect 11,000 carbon stars  in the LMC
and 2900 in the SMC. Groenewegen (\cite{groeniau}) calculated from the
Kontizas et  al. (\cite{kontdap}) survey  of the LMC, that  the carbon
stars amount  to $7750$.  Rebeirot et al.  (\cite{rebe}) on  the other
hand found $1707$ carbon stars in  the SMC. These numbers are not very
different  from our  estimated number  of  carbon stars  based on  the
colour--magnitude selection criteria.  We found, in the LMC, only 7572
C--type stars,  and we  think that we  are complete, except  for about
10\% misclassifications  (Cioni et  al. \cite{cionivar}) among  the M0
and M1  giants.  This  would increase the  number of  possible C--type
stars  detection   to  $10094$  and   $2243$  in  the  LMC   and  SMC,
respectively, and the present C/M ratio would be thus underestimated.  Other
missing C--type stars are those  with luminosities below the TRGB that
might have weaker CN bands and  may therefore have been missed as well
on the red grism plates used by Blanco \& McCarthy (\cite{bm}).

\subsection{The distribution of the C/M ratio over the face of LMC and SMC}

\begin{figure}
\resizebox{\hsize}{!}{\includegraphics{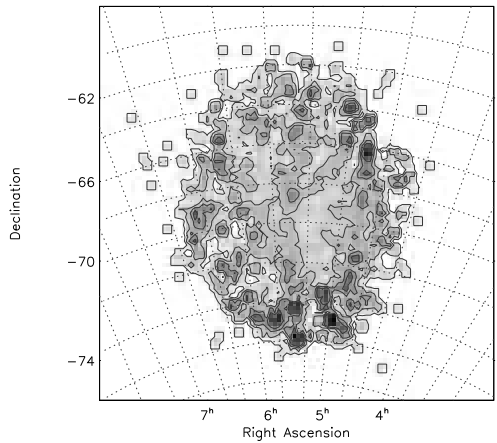}}
\caption{The C/M0+ ratio in the LMC. Contours are at: 0.1, 0.25, 0.4, 
  0.55. Darker regions correspond to a higher ratio.Coordinates
  correpond to the J2000 system.}
\label{lmcratio}
\end{figure}

\begin{figure}
\resizebox{\hsize}{!}{\includegraphics{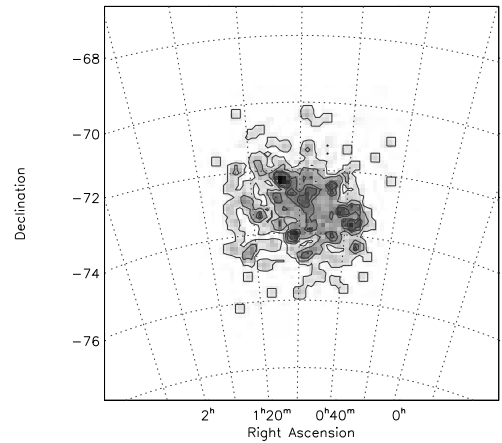}}
\caption{The C/M0+ ratio in the SMC. Contours are at: 0.1, 0.3, 0.5, 0.6. 
  Darker regions correspond to a higher ratio.}
\label{smcratio}
\end{figure}

From our samples of M--type and C--type AGB stars we have calculated
C/M0+ in all cells belonging to a grid a $100\times100$ of $0.04$
square degrees over the face of the LMC and the SMC. Fig.
\ref{lmcratio} and Fig. \ref{smcratio} show the ratios after boxcar
average smoothing of width 2. The most outstanding feature is the ring
of high C/M0+ values around the outer edge of the LMC.  In the SMC the
distribution is quite clumpy.  Both features are in contrast to what
is found for the distribution of all (M+C--type) AGB stars (Cioni et
al. \cite{cionimorf}): these all--AGB star distributions are smooth
and regular and they have the highest densities in the center of each
galaxy.

Are  the structures in Figures  \ref{lmcratio} and \ref{smcratio}
real?  We use $10^4$ squares and  a sample of 25,000 M--giants and 7000
C--stars, thus on average each bin  contains 2 M0+--type stars and 0.7 C
star, and  the ratio  C/M0+ is affected  by noise due  to small-number
statistics.   In the bin corresponding  to the highest ratio there are
3 C--  and 1 M0+-- star. In  bins where M--type stars  were not detected,
the C--type number of stars was  taken as an upper limit to the ratio,
i.e. a ratio  C/M0+=1 was assumed. To check  that the resulting pattern
does not  depend on the bin  size we also made  maps with $50\times50$
bins and  obtained the same  result. Now the  bin with the  lowest C/M
ratio contains  2 C-- and 4  M0+--type stars whereas the  bin with the
highest ratio contains 7 C-- and 7 M0+-stars. We conclude that the
variations seen in Figures  \ref{lmcratio} and \ref{smcratio} are real.

In  the LMC,  apart  from  the ring,  there  is a  small  region of  a
relatively  high C/M0+  value  at about  $\alpha=5^h30^m$ and  $\delta
=-67^\circ$. This structure points  approximately to the middle of the
bar and  defines two lobes  (NW \& SE)  of lower C/M0+ values  and two
other lobes (NE  \& SW) of higher C/M0+.  The  features may be related
to the effect  of the bar on the star formation  history or vice versa
as we do not know precisely when the bar has formed.  Considering that
the  metal--poor component is  probably old,  the bar  appears to  be a
fairly  recent  structure.  The  LMC bar  extends  approximately  from
$\alpha=5^h40^m$,     $\delta=-70^{\circ}$     to    $\alpha=5^h00^m$,
$\delta=-69^{\circ}$.

In the SMC the situation is more complex -- see Fig. \ref{smcratio}.
The fluctuations in this figure are real: the highest C/M ratio is
reached in a bin with 4 C--stars and 1 M0+--star and the lowest ratio in
a bin with 16 C-- and 16 M0+--stars.  The blob--like regions of high
C/M0+ do not form a regular pattern; there is a hint that C/M0+ is
higher in the external part of the central region.  The lowest ratio
is obtained in the so called ``wing''.  Probably the extension of the
SMC along the line of sight contributes to the irregular distribution.

Around the centre of the SMC the C/M0+ ratio is higher than inside the
ring of the LMC.  This is in agreement with earlier studies (e.g.
Blanco et al. \cite{bmb}) that show the metallicity of the AGB
population in the SMC to be lower than in the LMC.

\begin{table*}
\caption{C/M ratio in the LMC per ring}
\label{tabcmring}
\[
\begin{array}{clcccccc}
\hline
\noalign{\smallskip}
\mathrm{N} & \mathrm{Ring} & \mathrm{C} & \mathrm{C_{red}} 
& \mathrm{M0+} & \mathrm{M6+} & \mathrm{C/M0+} & \mathrm{C/M6+}\\
& & & & & \times\frac{{{A_{\mathrm RING}}}}{A_{\mathrm TOT}} 
& \times\frac{{{A_{\mathrm RING}}}}{A_{\mathrm TOT}} \\ 
(0) & \rho<2.5^{\circ} & 3906 & 1668 & 12071 & 2236 & 0.045 & 0.104\\
(1) & 2.5^{\circ}<\rho<3.4^{\circ} & 1220 &  572 &  3836 &  734 & 0.038 
& 0.092\\
(2) & 3.4^{\circ}<\rho<4.4^{\circ} & 1093 &  490 &  3288 &  586 & 0.058 
& 0.145\\
(3) & 4.4^{\circ}<\rho<5.5^{\circ} &  832 &  398 &  2824 &  538 & 0.071 
& 0.179\\
(4) & 5.5^{\circ}<\rho<6.7^{\circ} &  417 &  192 &  1881 &  271 & 0.072 
& 0.231\\
\noalign{\smallskip}
\hline
\end{array}
\]
\end{table*} 

\subsection{Related previous studies}

In Fig.  \ref{agb} most  of  the stars  populating the  (almost)
vertical  sequence  to the  left  of  the  line $(J-K_s)=1.4$  are  of
spectral  type  M0+. Blanco  \&  McCarthy  (\cite{bm}) discussed  the
ratios C/M2+,  C/M5+ and  C/M6+ and found  $0.2 (0.63)$,  $0.8 (4.3)$,
$2.2 (13.8)$  for the entire  LMC (SMC): the increase  illustrates the
existence of a large number of M--giants of early spectral type. In the
earlier  papers  (e.g.  Blanco,  McCarthy \&  Blanco  \cite{bmb})  the
samples are  incomplete in these earlier  types except in  a few small
fields mentioned  below: M--giants of  earlier spectral type  have weak
TiO and VO molecular bands and require spectra of very good quality to
derive the  subtype.  Blanco \&  McCarthy (\cite{bm}) stated  that the
estimated  number of  M2--M4  stars may  be  in error  by $20$\%.  The
authors showed that an  approximate linear relation exists between the
frequencies  of carbon  stars and  those  of M2--M4  giants: for  each
carbon star there are $4.6$ and  $1.58$ M2--M4 stars in the LMC and in
the SMC,  respectively. Westerlund (\cite{westerl})  suggests that the
most  massive AGB stars  could be  of early  type. These  stars should
exist in the Blanco fields as the immediate precursors of the detected
Cepheid population.

Frogel \& Blanco (\cite{frog}) detected stars with spectral types from
M0 to M5 but only in about 0.03 square degrees in the Bar West field
of the LMC. In the northern LMC Reid \& Mould (\cite{reidm})
spectroscopically identified O--rich stars of M0+ but only in six
small fields (total area of 0.8 square degrees in the northern
LMC.  They derive C/non--C ratios between  $0.19$ to $0.56$. The
similarity with the range of values obtained from Fig. \ref{lmcratio}
proves the validity of our photometric selection criteria for M0+
stars and C stars.

 Loup  et al.  (\cite{loup}) conclude that  in the  region roughly
indicated by $(J-K_s)\approx 1.1$ and  $(I-J)< 2.0 $ early M--type AGB
stars are mixed with C--rich AGB (Fig.~\ref{colcol}).  If we apply the
selection criteria given  by the straight lines in Fig. \ref{colcol} we
obtain  4456  C--type  and  765   M7+--type  AGBs  in  the  LMC.  The
distribution  of  the  C/M7+   ratio  is  strongly  dominated  by  the
distribution of  C--type stars and the  value of the ratio  is high in
the bar and progressively decreases  to the outer parts.  Using a 
coarser grid  ($50\times50$ instead of $100\times100$) as
discussed in Sect. 2.5) we  find a higher C/M7+ ratio outside
of the bar. In  both cases the numbers of stars are  low and the ratio
can be calculated  with confidence only in the inner  part of the disk
that is inside the metal  poor ring identified in Fig. \ref{lmcratio},
thus closer  to the bar.  Above we  have stated that AGB  stars of the
same luminosity but of lower metallicity have earlier spectra and this
is further supporting our conclusion that the stars in the ring in the
LMC have lower metallicity.

\section{A metallicity gradient in the LMC?}

The variation of the C/M0+--ratio may be caused by two effects: a
variation in metallicity or a variation in the star formation rate
depending on stellar mass. This second possibility will be discussed
in a later paper; here we assume that the metallicity variation
is the main cause of the C/M0+ variation.

To investigate in more detail the variation of the (C/M0+)--ratio over
the face of the LMC we calculated this ratio within segments of
concentric rings as follows (the same procedure was adopted in van der
Marel \& Cioni, \cite{structure}): the centre is at ($\alpha=5^h29^m$,
$\delta=-69.5^{\circ}$); the inner circle (i) has a radius of
$2.5^{\circ}$, and the rings are (ii) between $2.5^{\circ}$ and
$3.4^{\circ}$, (iii) between $3.4^{\circ}$ and $4.4^{\circ}$, (iv)
between $4.4^{\circ}$ and $5.5^{\circ}$ and (v) between $5.5^{\circ}$
and $6.7^{\circ}$.  Table \ref{tabcmring} lists: (i) the number of the
ring, (ii) inner and outer limit of the ring, (iii) the number of
C--type and M--type stars as obtained from the global selection (C \&
M0+); (iv) similar numbers from a sample using the more restrictive
selection criteria derived by Loup et al. (e.g.~Sect.~2.3), (v) the
average of C/M0+ and C/M6+ multiplied by the fractional area covered
by each ring ($A_{\mathrm RING}/A_{\mathrm TOT}$).  The average value
of the surface density increases for larger distance from the center;
excluding the very central region, the trend deduced from Fig.
\ref{lmcratio} is conserved.

\begin{figure}
\resizebox{\hsize}{!}{\includegraphics{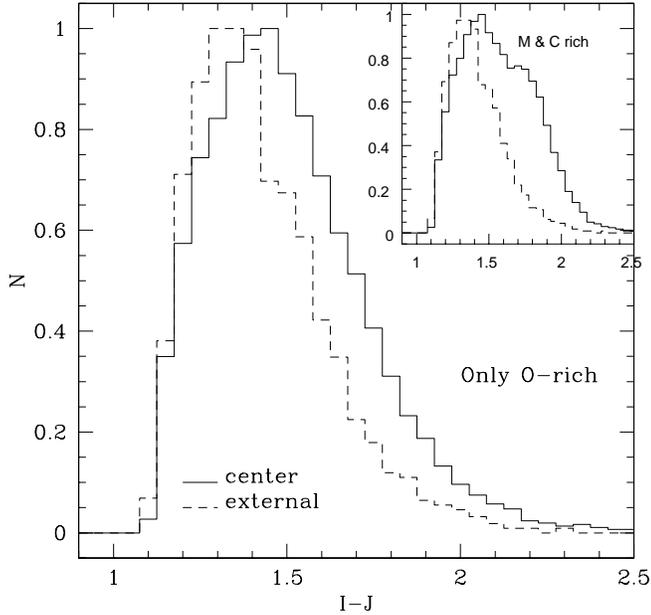}}
\caption{Histograms of the distribution of the $(I-J)$ colour of 
  sources in the central part (continuous line) and in the most
  external ring (dashed line) of the LMC. Regions are defined in the
  text. The central diagram includes only O--rich sources and the
  diagram on the upper right corner includes as well C--rich sources.}
\label{ijcol}
\end{figure}

  If the  outermost  ring  marks lower  metallicity  it must  also
contain relatively more early M--giants than the central area. This is
indeed  the case:  see  Fig. \ref{ijcol},  that  shows the  normalised
number of O--rich sources as a function of $(I-J)$ colour in the central
($\rho<2.5^{\circ}$)      and      in      the     outermost      ring
($5.5^{\circ}<\rho<6.7^{\circ}$) of  the LMC.  The panel on  the upper
right corner shows  the same histograms with C--type  stars added; the
latter produce a bump in  the distribution at about $(I-J)=1.7$.  In the
outermost  ring  the  distribution  of  the  colours  peaks  at  about
$(I-J)=1.3$ and the  histogram of the colours of  the central area peaks
at  $1.5$.  At about  $(I-J)=1.7$ (M5)  there are  almost twice  as many
sources in the  center as in the outer ring.  This  is the effect that
we predicted. Unfortunately we cannot exclude the possibility that the
shift of the peak of the histograms is due to exinction.  If we assume
that the  reddening decreases linearly  with increasing radius  with a
step of $(I-J)=0.025$ in the range  $0.1-0$ from the center to the outer
regions, and  we recalculate the  C/M ratio we obtain  $0.33$, $0.59$,
$2.94$   and  $38.4$,  for   the  C/M0+,   C/M3+,  C/M5+   and  C/M7+,
respectively. In  addition there are  sources in the innermost  bin at
$(I-J)_0<1.02$ which would imply a spectral type earlier than M0. Thus
we conclude  that if a  dif\-fe\-ren\-tial reddening between the  center and
the outskirts of  the LMC is taken into account,  giants in the center
would have, on average, an earlier spectral type than giants in the outskirts.

\begin{figure}
\resizebox{\hsize}{!}{\includegraphics{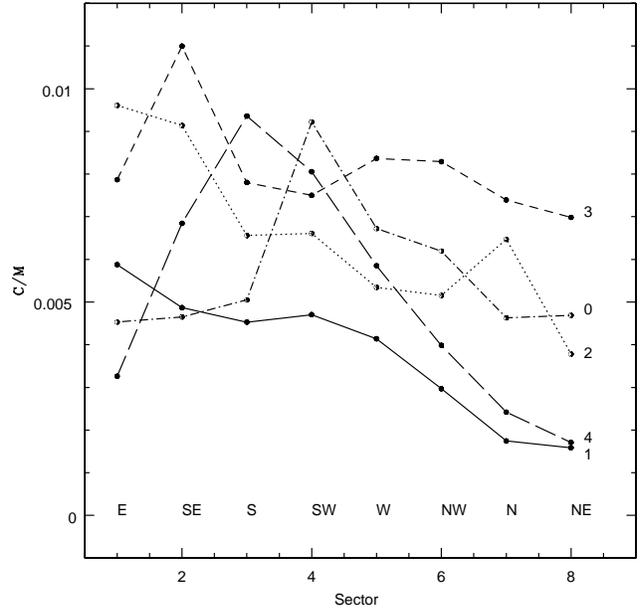}}
\caption{The C/M ratio in the LMC, multiplied by the fractional area 
  of the circular corona within which it has been calculated, is shown
  as a function of sector in each ring. Sector numbers correspond to
  increasing angle. The cardinal points indicate the starting
  direction of each sector.  Each line represents the ratio in a ring:
  $\rho<2.5^{\circ}$ (0 -- dot--short--dashed line),
  $2.5^{\circ}<\rho<3.4^{\circ}$ (1 -- continuous line),
  $3.4^{\circ}<\rho<4.4^{\circ}$ (2 -- dotted line),
  $4.4^{\circ}<\rho<5.5^{\circ}$ (3 -- short--dashed line) and
  $5.5^{\circ}<\rho<6.7^{\circ}$ (4 -- long--dashed line),
  respectively.}
\label{cmring}
\end{figure}

The next step is to divide each ring in the LMC into eight segments of
$45^\circ$ and labelled  counterclockwise from 1 through 8  or by wind
directions (E, SE, S, SW, W,  NW, N and NE) per increasing angle.  The
first sector starts in the East and ends in the North--East direction.
Fig. \ref{cmring} shows  the C/M0+  ratio averaged over  each segment.
On average  the C/M0+ ratio reaches  it highest value  in the southern
segment (number  3 or direction  S; see Fig. \ref{cmring}), that  is in
the same direction as the HI  ``bridge'' that connects the LMC and the
SMC. It thus seems that we  have detected an increase of the C/M ratio
towards  the  bridge  connecting  the  two Clouds  and  thus  a  lower
metallicity. A lower  metallicity has also been proposed  by Muller et
al. (\cite{muller})  because of the  detection of  a $^{12}$CO(1--0)
emission  region within  the Magellanic  Bridge ($\alpha=1^h56^m47^s$,
$\delta=-74^{\circ}17^{\prime}41^{\prime\prime}$).

\begin{figure}
\resizebox{\hsize}{!}{\includegraphics{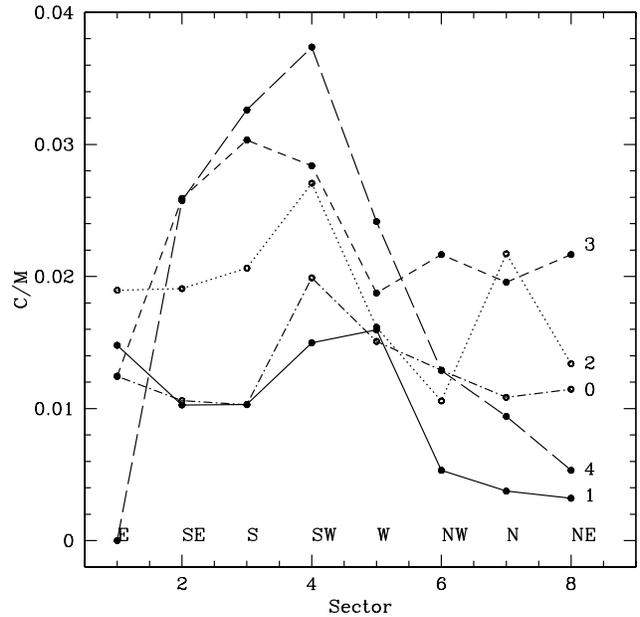}}
\caption{Same as Fig. 5 but with a different selection of stars (see text).}
\label{cmring1}
\end{figure}

The same procedure has been applied to the latest type AGB stars
selected according to the Loup et al.  (\cite{loup}) criteria; the
result is shown in Fig. \ref{cmring1}.  In each ring the C/M ratio has
a peak.  {The innermost and the outermost ring have a peak in the SW
  region, while in the other rings the peak moves from W to S going
  from inside to outside the LMC.}  Note that the ratio C/M is higher
in Fig.  \ref{cmring1} than in Fig. \ref{cmring}, because we excluded
early--type M--giants.  The average C/M ratio over the whole LMC
($0.19\times A_{\mathrm TOT}$) is about $2.68$ which is very close to
the value of $2.2\pm0.1$ obtained by Blanco \& McCarthy (\cite{bm})
selecting carbon stars and giants of M6+ spectroscopically.
 
\subsection{Calibration of C/M in terms of [Fe/H]}

In this section we quantify the  C/M ratio in terms of [Fe/H] by using
measurements  available in the  literature for  galaxies in  the Local
Group with different metallicity.

\begin{figure}
\resizebox{\hsize}{!}{\includegraphics{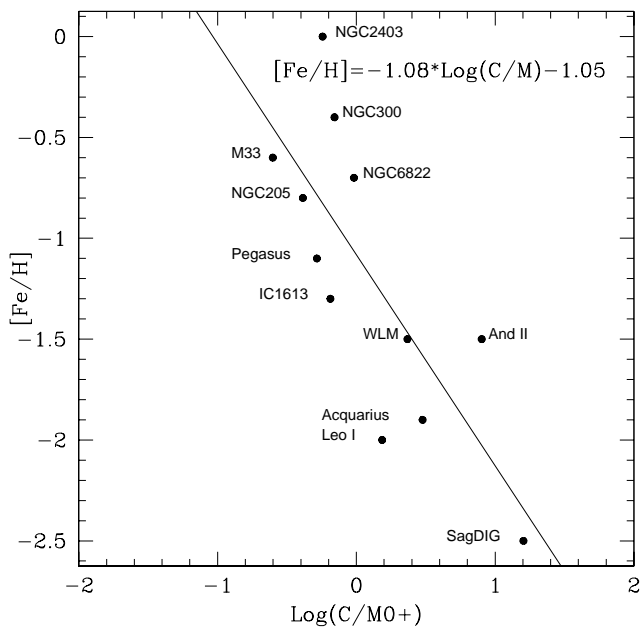}}
\caption{Relation between the metallicity ([Fe/H]) and the decimal 
logarithm of the C/M ratio for M0+ stars. Data are taken from 
Groenewegen (\cite{groeniau}). The solid line is a linear fit through 
the data.}
\label{metal}
\end{figure}

We took as  a reference the work by  Groenewegen (\cite{groeniau}). He
lists the metallicity attributed to  most of the galaxies in the Local
Group  with the  number of  carbon-- and  oxygen--rich AGBs  stars. We
considered only those galaxies for which AGBs were detected (O and C).
Fig. \ref{metal} shows  the decimal logarithm of the  C/M ratio versus
metallicity.   There  is  a  clear  correlation  and  the  solid  line
indicates the best fit through the points.

If  we use the  same relation  to quantify  the spread  in metallicity
derived in the previous sections  for the Magellanic Clouds, using the
values corresponding to the lower and higher contours, we obtain about
$0.75$  dex between  the central  area and  the outermost  ring.  This
value is in agreement with the spread inferred from the me\-tal\-li\-ci\-ty of
the star clusters in the LMC (Kontizas et al. \cite{konti}).  The
error on  the linear fit  is about $0.1$  dex and we believe  that the
error on the  C/M ratio is negligible due to the  large number of AGBs
involved  and to  the completeness  of detection.  However,  the ratio
depends  on the  selection criteria  used (cf.  Table \ref{tabcmring},
cols. 6--7).   Mouhcine \& Lan\c{c}on (\cite{mouci})  have shown that
different  star  formation histories  lead  to  non--linear curves  in
diagrams such as Fig. \ref{metal},  but for populations older than $1$
Gyr the models merge into a  common sequence.  Here, we are not giving
an absolute value for the metallicities in the Clouds because with the
photometric  data available  at  the moment  we  cannot determine  the
zero--point of the C/M versus metallicity relation. The  
metallicity spread detected in the MCs agrees with the spread measured  
in Local Group dwarf galaxies (i.g. Shetrone et al. \cite{shet1}, 
\cite{shet3}, Tolstoy et al. \cite{eline}). 
However, the spread in metallicity
 may influence the fit in Fig. \ref{metal}. 

\subsection{Previous studies on abundance gradients in the MCs}
The indication of a metallicity gradient  in the LMC and of a range of
metallicities  in  the SMC  obtained  in this  work  is  based on  two
fundaments:  (i)  our survey  of  the MCs  is  complete  and (ii)  our
photometry allows us a statistical identification of the AGB stars of
different  atmospheric   abundances.   Previous  studies   of  varying
metallicity  have  been quite  localised.   For  example, Nikolaev  \&
Weinberg (\cite{nw}) do not find an indication of a radial metallicity
gradient in  the LMC using isochrone  fitting to the  red giant branch
located  in  a   centre ($\alpha=5^h24^m$,  $\delta=-69.8^{\circ}$)  
and  in  an  outer ($\alpha=6^h12^m$, $\delta=-67.5^{\circ}$)  field   
of  the  LMC,  of $2^{\circ}\times1^{\circ}$  in  size.    
 Comparing the  location of  these fields with  Fig.\ref{lmcratio}, our
data  confirm  that  between  these  fields there  is  no  significant
difference in abundances.

The most  extensive work  done in  the eighties is  that of  Blanco \&
McCarthy  (\cite{bm}). They surveyed  the SMC  and the  LMC down  to a
limiting magnitude of $I=17$ in $37$ and $52$ fields, respectively, of
$0.12$  square degrees  each.   Their survey  is  $95$\% complete  for
C--type and M5+--type AGBs while the numbers of M2--M4 stars may be in
error by as much  as $20$\%. They found that the C/M  ratio in the LMC
does not vary  appreciably away from the central  regions while in the
SMC  the ratio  may  decrease towards  the  peripheral regions.   They
further  investigated the distribution  of the  ratios by  binning the
number of AGBs C-- and  O--rich in three groups: central, intermediate
and outer.  They  concluded that there is no  apparent gradient in the
LMC while in  the SMC a lower ratio in the  outer field was confirmed.
Note that these  ratios include O--rich AGB stars of  M6+ and that the
variation of the C/M ratio in  the SMC disappears if also M2--M4 stars
are included.   The most important thing  to notice is  that their LMC
fields  are  limited  to  $-65^{\circ}<\delta  <-74^{\circ}$  and  the
highest  value  of   the  C/M  ratio  that  we   detect  is  at  about
$\delta=-75^{\circ}$    (Fig. \ref{lmcratio}).     Reid    \&    Mould
(\cite{reidm})  took spectra of  all the  stars in  six fields  in the
northern  part  of  the  LMC,  three  of which  contain  a  Blanco  et
al.  (\cite{bmb}) field.  Reid \&  Mould  concluded that  if a  radial
variation of  the C/M ratio exists,  then the Bar is  less enriched in
heavy elements than the outer regions. Comparing the location of their
analysed fields  with Fig. \ref{lmcratio},  we find that  these fields
are not representative of the radial distribution.

A positive indication of a metallicity spread in the Magellanic Clouds
comes  from   Luck  \&  Lambert  (\cite{lula})   who  concluded,  from
measurements in Cepheids and non--variable supergiants that the [Fe/H]
ratio  in the  SMC has  an  abundance dispersion  consistent with  the
errors while  in the LMC a  real dispersion is  possible.  In addition
Spite   et   al.    (\cite{spita}a,  \cite{spitb}b)   suggested   that
inhomogeneities in  the abundances derived  from F and G  super giants
are  due  to  a  low  metallicity  in the  halo  and  a  progressively
increasing metallicity in the disk.

Kontizas et al. (\cite{konti}) studied  a sample of $94$ star clusters
in  the LMC of  intermediate age  (between $0.5$  and $3$  Gyr).  They
found that the outermost  clusters are significantly metal poorer than
those at  the centre. Their  outer system coincides with  our outermost
region (see their Fig.~1) where we detect the highest C/M ratios.

\subsection{Metallicity gradients in other galaxies of the Local Group}

There is evidence, though somewhat controversial, that the metallicity
is a decreasing function of radius in other galaxies in the Local
Group.

In our Galaxy the metallicity gradient is present among clusters
(Friel \cite{fri}).  Note that clusters do not u\-su\-al\-ly contain C
stars even if they are metal poor. This is strictly an age effect and
the C/M is directly correlated with the intermediate age stars.
Andrievsky et al. \cite{andrea} recently concluded that a linear fit
to the [Fe/H] data of Galactic Cepheids from $4$ to $12$ kpc produces
a grandient of $-0.06$ dex kpc$^{-1}$.

Fields in the halo and in the centre of M31 were observed in $B$ and
$V$, the colour variation corresponds to about $0.2$ dex in [Fe/H]
(Van den Bergh \& Pritchet \cite{vdbpri}).  In M31 Brewer et al.
(\cite{bre}) found an increase of the C/M ratio with galactocentric
distance by studying seven fields spaced along the M31's SW
semi--major axis. On the other hand, Freedman \& Madore
(\cite{fremad}) argue that because the period--luminosity relation of
Cepheids does not vary with radius in M31 there is no abundance
gradient.  Globular clusters in M31 show a spread in metallicity
within $5$ kpc from the centre but metal rich clusters exist in the
outer halo (Huchra et al. \cite{huc}).

Nitrogen and oxygen lements present a strong radial gradient in M33
(Smith et al.  \cite{smithal}).In IC 1613 a slight C/M gradient from
the center outward has been detected.  Stars of the two spectral types
were selected using the (CN-TiO$)/(R-I)$ diagram. M0+ stars were also
included (Albert et al.~\cite{albert}).
  
In the dwarf spheroidals NGC147, NGC187 and NGC205 the width of the
RGB is the major indication of a metallicity spread which is often
larger in the centre of the galaxies than in the outer parts (Han et
al.  \cite{han}, Mart\'{i}nez--Delgado \& Aparicio \cite{mda}, Mould
et al.  \cite{mkdc}).  The Sculptor galaxy presents a strong radial
population effect (Hurley--Keller et al. \cite{hkmg}) and it would be
interesting to know if and how this is associated to a gradient in
metallicity.

\section{Conclusions}
We have derived C/M ratios in the Magellanic Clouds from AGB stars
selected in colour--magnitude diagrams obtained from the DCMC
catalogue.  This value is a simple, but robust indication of the
metallicity of the AGB population. It varies considerably across the
surface of both galaxies.  In the LMC we find a increase of the ratio
C/M0+ in at the inside of the elliptical distribution of the AGB
star. The distribution of the C/M valueis, however, are rather clumpy
and that has prevented previous authors from detecting such a gradient. A
more detailed analysis of the LMC C/M ratio shows that towards the
direction of the bridge connecting the two Clouds the ratio C/M0+
is still higher.

The range of variation of the C/M ratio has been ca\-li\-bra\-ted in
terms of [Fe/H]; in the LMC it corresponds to a spread in metallicity
of $0.75$ dex. The same spread has been found in the SMC and it
approximately corresponds to the range of values derived in
intermediate age LMC star clusters. The SMC has on average a lower
metallicity than the LMC and the distribution of regions of high C/M0+
shows no systemnatic behaviour. 

To conclude, it is important to remember that the variation of
metallicity found is valid for the epoch when these giants formed and
it is not ne\-ces\-sa\-ri\-ly valid today or for other epochs better
traced by other stellar populations or gas. On the other hand in view
of the recent results by Mouhcine \& Lan\c{c}on (\cite{mouci}) the C/M
ratio measures both the metallicity of the carbon star progenitors and
the present interstellar medium metallicity, unless the local
composition has been strongly altered by the presence of a Bar or
other interactions.

\begin{acknowledgements}
  The authors thank Martino Romaniello, Luca Pasquini and Vanessa Hill
  for useful discussions during the preparation of this article.
\end{acknowledgements}

\end{document}